\documentclass[showpacs,superscriptaddress,floatfix,eqsecnum,aps,nofootinbib]{revtex4}
\usepackage{epsfig}

\usepackage{graphicx}
\usepackage{amsmath, amssymb}

\begin{document}

\title{$SU(5)$ grand unification on a domain-wall brane from an $E_6$-invariant
action}

\author{Aharon Davidson}\email{davidson@bgu.ac.il}
\affiliation{Physics Department, Ben-Gurion University, Beer-Sheva 84105, Israel}
\author{Damien P. George}\email{d.george@physics.unimelb.edu.au}
\affiliation{School of Physics, Research Centre for High Energy
Physics, The University of Melbourne, Victoria 3010, Australia}
\author{Archil Kobakhidze}\email{archilk@unimelb.edu.au}
\affiliation{School of Physics, Research Centre for High Energy
Physics, The University of Melbourne, Victoria 3010, Australia}
\author{Raymond R. Volkas}\email{raymondv@unimelb.edu.au}
\affiliation{School of Physics, Research Centre for High Energy
Physics, The University of Melbourne, Victoria 3010, Australia}
\author{Kameshwar C. Wali}\email{wali@phy.syr.edu}
\affiliation{Physics Department, Syracuse University, Syracuse NY 13244-1130, U.S.A.}

\begin{abstract}
An $SU(5)$ grand unification scheme for effective $3+1$-dimensional fields 
dynamically localised on a domain-wall brane is constructed.  This is
achieved through the confluence of the clash-of-symmetries mechanism for 
symmetry breaking through domain-wall formation, and the Dvali-Shifman 
gauge-boson localisation idea.  It requires an $E_6$ gauge-invariant
action, yielding a domain-wall solution that has $E_6$ broken to
differently embedded $SO(10) \otimes U(1)$ subgroups in the two bulk
regions on opposite sides of the wall.  On the wall itself, the unbroken
symmetry is the intersection of the two bulk subgroups, and contains $SU(5)$.
A $4+1$-dimensional fermion family in the $27$ of $E_6$ gives rise to
localised left-handed zero-modes in the $5^* \oplus 10 \oplus 1 \oplus 1$
representation of $SU(5)$.  The remaining ten fermion components of the $27$
are delocalised exotic states, not appearing in the effective $3+1$-dimensional
theory on the domain-wall brane.  The scheme is compatible with the type-2
Randall-Sundrum mechanism for graviton localisation; the single extra dimension
is infinite.
\end{abstract}

%\pacs{}

\maketitle

\section{Introduction}

If our universe is a $3+1$-dimensional brane 
\cite{Rub&Shap, Akama, Visser, GibbonsWiltshire, ADD, Antoniadis, Antoniadisetal, RS1, RS2}
existing in a $4+1$-dimensional
spacetime, then the most likely field-theoretic origin for the brane is a
scalar-field domain wall (DW) or kink \cite{Rub&Shap}.  This generic idea is naturally compatible
with the type-2 Randall-Sundrum (RS2) mechanism for producing effective $3+1$-d
gravity on the brane \cite{RS2} (see, for example, Refs.~\cite{Gremm1, DeWolfe, DavidsonMannheim}
for the extension of thin-brane RS2 to a domain-wall brane).  The challenge is to dynamically localise all the other
ingredients necessary for a phenomenologically successful effective theory
on the brane: gauge bosons, fermions, and Higgs bosons.  Various localisation
ideas for these disparate classes of fields have been recently combined to
produce an effective brane theory that is plausibly very similar to the
standard model \cite{DGV}.

The purpose of this paper is twofold.  First, we wish to point out a very
elegant generic connection between the clash-of-symmetries (CoS) mechanism for symmetry
breaking through domain wall formation \cite{DavidsonClash1, Rozowsky, Volkasclash, Shin}\footnote{See also
\cite{PogosianVachaspati1, Vachaspati, PogosianVachaspati2} for related works,
and \cite{DSsusy} for soliton-induced supersymmetry breaking.}, 
and the Dvali-Shifman (DS) idea for
dynamical gauge-boson localisation \cite{DvaliShifman}.  Second, we use this remarkable confluence
to construct an explicit scheme that realises an $SU(5)$ gauge-invariant
effective theory on the brane.  In a sense, it is a grand unified extension
of the model of Ref.~\cite{DGV}, but the way in which the Dvali-Shifman mechanism is
realised is quite different, and we shall argue that it is in fact
conceptually more advanced.  Remarkably, this scheme immediately produces
a realistic spectrum of localised fermion zero-modes \cite{Jackiw&Rebbi} (for a review see
\cite{rubrev}) using the simplest possible mechanism.  While it is beyond the scope of this paper
to write down a complete phenomenologically-acceptable domain-wall-brane localised $SU(5)$ theory, we
shall conclude with brief remarks about how this could be attempted. 

The clash-of-symmetries phenomenon 
\cite{DavidsonClash1, Rozowsky, Volkasclash, Shin, PogosianVachaspati1, Vachaspati, PogosianVachaspati2}
automatically arises when the simple
$Z_2$ kink is extended to a theory with a continuous internal symmetry group $G$
in addition to the discrete symmetry.  Taking the scalar-field multiplet to
be in a non-trivial representation of $G$, the domain-wall configuration spontaneously
breaks $G$ in addition to reflecting the disconnected vacuum manifold topology
created by the spontaneous breaking of the discrete symmetry.  Two classes of
domain-wall solutions exist: those which respect the same subgroup $H$ of $G$
at all values of the bulk coordinate $y$, and those where the unbroken subgroup varies
in the bulk.  We shall call the first class ``non-CoS domain walls'', contrasted
with the ``CoS domain walls'' of the second class.  Clash-of-symmetries DWs can arise
when the subgroups respected asymptotically (at $y = \pm \infty$) 
are isomorphic but {\em differently embedded} subgroups, $H$ and $H'$.  
The symmetry group at finite $y$ is typically the intersection $H \cap H'$, which
is of course smaller than both $H$ and $H'$.

The last observation provides an immediate connection with the Dvali-Shifman
proposal for dynamical gauge-boson localisation.  The DS mechanism, as originally
proposed \cite{DvaliShifman}, envisaged a domain wall configuration where the full group $G$ is
restored in the bulk, but broken to $H$ in the wall.  The gauge bosons of $H$
propagate on the wall either as massless Abelian gauge fields or glueballs formed
from non-Abelian gauge fields.
In the bulk, {\em all} gauge bosons have to be 
incorporated into massive $G$-glueballs.\footnote{The Dvali-Shifman mechanism requires a confining
$4+1$-dimensional gauge theory to live in the bulk.  The issue of confinement in $4+1$ dimensions
is not completely understood, so the Dvali-Shifman mechanism in that context has the status of being 
a plausible conjecture.  There is good lattice gauge theory evidence that pure $SU(2)$ Yang-Mills theory with
an ultraviolet cut-off is confining in $4+1$ dimensions when the gauge coupling constant exceeds a certain
critical value \cite{Creutz}.  It is thus plausible that a variety of $4+1$ dimensional gauge theories
exhibit confinement at sufficiently large coupling strength.}  
Thus, the massless Abelian gauge fields
on the wall have to become incorporated into massive glueballs in the bulk, and
the energy cost associated with the mass gap then plausibly localises them to the wall.
This heuristic argument is bolstered by the 
dual-superconductivity model \cite{tHooft_dual, Mandelstam} for the
confining bulk: the electric field lines from a source charge in the wall are repelled from
the interface with the dual-superconducting bulk \cite{A-H&Schmaltz2, RubaDub}, just as magnetic field lines
are Meissner-repelled from an ordinary superconductor.  The non-Abelian gauge fields
of $H$ are also plausibly localised if the mass of the $G$-glueballs exceeds the
mass of the $H$-glueballs. 

The fact that the full symmetry $G$ is asymptotically restored is clearly not a necessary
condition.  In the CoS situation, the brane-group $H \cap H'$ is a subgroup of both
$H$ and $H'$, the unbroken symmetries in the two semi-infinite bulk regions.  By the
DS reasoning, provided $H$ and $H'$ contain confining non-Abelian factors, at least
some of the gauge bosons of $H \cap H'$ will be localised.  For a realistic theory,
we need the localised gauge bosons to include those of the standard model.  The
model-builder needs to engineer the theory to achieve this effect.  While this engineering
shall be the main concern in the rest of the paper,
our first generic point has already been made: {\em  the clash-of-symmetries automatically
gives rise to Dvali-Shifman gauge-boson localisation}.  This CoS alternative realisation of
the DS mechanism seems conceptually neater than the original, because it can be achieved
using scalars in a single irreducible representation of $G$.  The original requires two
multiplets: a $G$-singlet to form a kink, which in turn forces a $G$-multiplet to condense
in the core of the wall.

We shall show that the CoS-DS confluence can naturally produce an $SU(5)$ effective theory
on the brane.  The basic ingredients are $G = E_6$, with the DW-producing scalar field
in the adjoint or $78$ representation.  The groups $H$ and $H'$ will be the 
differently-embedded maximal subgroups $SO(10) \otimes U(1)_E$ and $SO(10)' \otimes U(1)_{E'}$,
respectively.  Their intersection is $SU(5) \otimes U(1) \otimes U(1)$, with $SU(5)$ of
course being a subgroup of both $SO(10)$ and $SO(10)'$.  Taking both of those
as confining gauge theories in the bulk, the localisation of $SU(5)$ gauge bosons follows
from the DS-effect.  The gauge fields of $U(1) \otimes U(1)$ are {\em not} completely
localised.  When $4+1$-d fermions in the $27$ of $E_6$ are Yukawa-coupled to the scalar
multiplet, we shall show that $3+1$-d left-chiral zero-modes in the 
phenomenologically-realistic $5^* \otimes 10 \oplus 1
\oplus 1$ representation of $SU(5)$ are localised.  The remaining ten fermion components remain $4+1$-d,
and are thus absent from the effective brane-theory. 
The result that the chiralities of the zero-modes come out to be
phenomenologically correct is very non-trivial, as we shall explain.

We review the clash-of-symmetries idea in Sec.\ref{CoS}.  Section \ref{SO10}
describes a warm-up example featuring $SO(10)$ CoS domain walls, and explains
why the extension to $E_6$ constructed in Sec.\ref{E6} is needed.  We conclude
in Sec.\ref{conc}.

\section{The clash of symmetries}
\label{CoS}

Consider a theory (an action) whose symmetry group is the direct product of 
a continuous symmetry group $G$ and a discrete symmetry $Z$.  It is important that
$Z$ is {\em not} a subgroup of $G$.  
Suppose the global minima of the Higgs potential spontaneously break $G$ to subgroup $H$,
and simultaneously break $Z$ to a smaller discrete group.  For the sake of definiteness,
we shall take the $Z = Z_2 = \{1,z\ {\rm s.t.}\ z^2=1\}$ example in what follows, 
with the $Z_2$ completely broken.

The vacuum manifold then consists of two disconnected copies of the coset space $G/H$,
with the copies related by the spontaneously broken $z \in Z_2$.  This is an immediate
generalisation of the simple $Z_2$ kink situation, where the vacuum manifold consists of
just two disconnected points related by $Z_2$.  Each such point is expanded into the
non-trivial manifold $G/H$.  We shall call the disconnected pieces $G/H$ and $(G/H)_z$.
The $Z_2$ must not be a subgroup of $G$ for the two disconnected pieces to exist.

\begin{figure}
\centering
\includegraphics[width=0.5\textwidth]{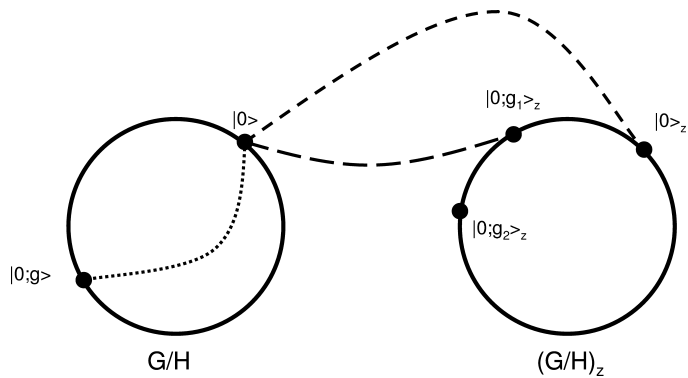}
  \caption{The vacuum manifold of a $G \otimes Z_2 \to H$ model.  The two circles schematically depict the
disconnected coset spaces $G/H$ and $(G/H)_z$.  Each point along the $G/H$ circle corresponds to
a vacuum $|0;g\rangle$ for some $g \in G$, with the corresponding situation for the $(G/H)_z$ circle.
The three broken lines represent possible domain wall configurations, with the endpoints at $y = \pm \infty$
on various choices of vacua.
The dotted line represents a possible non-topological domain wall configuration.  The short-dashed
line represents a non clash-of-symmetries domain wall configuration, while the long-dashed line
is a clash-of-symmetries domain wall.}
  \label{fig:GH}
\end{figure}

Let $|0\rangle$ be an element of $G/H$.  By definition, $h|0\rangle = |0\rangle$ for all
$h \in H$.  Since the Higgs potential is $G$-invariant, if we apply a transformation
$g \in G/H$ (that is, a transformation such that $g \in G$ but $g \not\in H$) to $|0\rangle$,
we obtain a degenerate vacuum state $|0;g\rangle \equiv g|0\rangle$.  By considering all
possible $g$'s, these transformations generate the $G/H$ piece of the vacuum manifold.
Applying the non-identity transformation $z \in Z_2$ from the discrete group, we obtain
the discrete image  $|0\rangle_z \equiv z|0\rangle$ of $|0\rangle$.  This image is
a point in the other disconnected piece $(G/H)_z$ of the full vacuum manifold.  By applying
all possible $g \in G/H$ to $|0\rangle_z$, the space $(G/H)_z$ is generated.  Figure \ref{fig:GH}
illustrates this situation.

The degenerate vacua $|0\rangle$ and $|0;g\rangle$ respect {\em differently embedded}
but otherwise isomorphic subgroups $H$ and $H_g$, respectively.  This is elementary:
Let $h_1, h_2 \in H$ such that $h_1 h_2 = h_3 \in H$.  Then the conjugates $g h_{1,2,3} g^{-1}$
respect the same multiplication table and hence the set $g H g^{-1}$ is precisely $H_g$
which is isomorphic to $H$ but a different subset of $G$.  If $h |0\rangle = |0\rangle$,
then trivially $g h g^{-1} g |0\rangle = g |0\rangle$, which simply says that the
conjugated elements preserve the other vacuum: $(g h g^{-1}) |0;g\rangle = |0;g\rangle$.
Similar statements are true for $(G/H)_z$.

The boundary conditions at $y = \pm \infty$ for domain wall configurations are chosen 
from the vacua.  If the chosen vacua are either both from $G/H$, or both from $(G/H)_z$,
then the ``domain wall'' configurations are not topologically stable:  they are in the
same topological class as any of the spatially-homogeneous vacua $|0;g\rangle$, 
or respectively $|0;g\rangle_z$, and will
dynamically decay to one of these vacua.  They may be metastable, depending on the
Higgs potential topography,\footnote{We dread to use the term ``landscape''.} so
while they are of some interest we shall not consider them further in this paper.

Topologically non-trivial DW configurations have one boundary condition from $G/H$
and the other from $(G/H)_z$.  Evidently, there is an uncountable infinity of such
choices, and thus potentially an uncountable infinity of DW solutions, all within the
same non-trivial topological class.  Figure \ref{fig:GH} illustrates the plethora of choices.
This potential richness has no analogue for the simple $Z_2$ kink.

Suppose that the boundary condition at $y = -\infty$ is $|0\rangle$ and at $y = +\infty$
it is $|0\rangle_z$.  Then if $h|0\rangle = |0\rangle$, it also follows that
\begin{equation}
h|0\rangle_z \equiv h z |0\rangle = z h |0\rangle = z |0\rangle \equiv |0\rangle_z,
\label{eq:hz}
\end{equation}
because by assumption the symmetry is $G \otimes Z$ so that $g z = z g \forall z \in Z,g \in G$ 
and hence $h z = z h$ always.  Thus, the unbroken symmetry at $y = -\infty$ is precisely
the same set $H$ as at $y = +\infty$.  A domain wall configuration that interpolates
between these vacua is then expected to respect the same subgroup $H$ at all $y$.
This is an example of a non-CoS domain wall.  Clearly, taking the vacua as
any pair $|0;g\rangle$ and $|0;g\rangle_z$ produces a similar outcome (the resulting
configuration is nothing more than the $g$ transform of the original one).  A non-CoS
domain wall is the simplest possible generalisation of a $Z_2$ kink for a $G$-invariant
theory.

However, there is obviously a second, more interesting possibility:  if the vacuum is
$|0\rangle$ at $y = -\infty$, then the vacuum at $y = +\infty$ can also be a $|0;g\rangle_z$
for $g \neq 1$.  In that case, the subgroups respected asymptotically are the 
differently-embedded but isomorphic groups $H$ and $H_g$, respectively.  This defines
a CoS-style domain 
wall \cite{DavidsonClash1, Rozowsky, Volkasclash, Shin, PogosianVachaspati1, Vachaspati, PogosianVachaspati2}.  
At finite $y$, the configuration would be expected to
respect the smaller group $H \cap H_g$ due to the fact that the solution has
to ``reconcile'' boundary conditions that have different stability groups 
that ``clash''.\footnote{From experience, we have found that $H \cap H_g$ is the
usual outcome.  The specifics depend on the case considered.
Sometimes there is enhanced symmetry at $y=0$ because some of the 
scalar multiplet components instantaneously vanish there.  This enhancement on
a set of measure zero has yet to find application, although speculations exist \cite{Shin, DemariaVolkas}.}

So, there will be an infinite family of non-CoS DWs, trivially related to each other
by global transformations $g \in G/H$.  They all have the same energy density, because the
Hamiltonian is invariant under $G$.  The CoS DWs have a more complicated spectrum.
Consider two configurations, $\chi_1(y)$ and $\chi_2(y)$,
with $\chi_1$ interpolating between $|0\rangle$ and $|0;g_1\rangle_z$, while
$\chi_2$ interpolates between $|0\rangle$ and $|0;g_2\rangle_z$, such that $g_1 \neq g_2$.
Suppose, for the moment, that $G$ is a global but not a local symmetry.  These two
solutions {\em cannot} be transformed into each by a {\em global} $G$-transformation,
so they would be expected to have different energy densities (their configurations
trace different paths through the Higgs potential topography).  As a corollary, the non-CoS
solutions should have a different energy density from the CoS solutions.  All these
solutions are in the same topological class, so finite-energy dynamical evolution
between them is allowed.  Hence, the special configurations within that
topological class that minimise the energy density will be topologically stable.
The others should be unstable to decay to the minimum-energy configurations, which
play the role of ``vacua'' for the ``kink-sector''.  This general reasoning cannot
tell you which configuration has the minimum energy-density:  you need to calculate
that within a specific model.  For example, in the toy model 
considered in Ref.~\cite{DavidsonClash1} the sign of
a Higgs potential parameter determined whether the non-CoS or a CoS solution
was energetically favoured. 

Suppose now that $G$ is a gauge symmetry, and again consider the configurations
discussed in the previous paragraph, together with the specification of 
vanishing gauge fields $A_M$ at the solution-level.  
The non-CoS solutions remain connected through global transformations.  Two CoS
scalar field configurations, $\chi_1$ and $\chi_2$, can be written as local $G$-transforms
of each other.  Suppose that
\begin{equation}
\chi_2(y) = U(y) \chi_1(y),
\label{eq:chi1chi2}
\end{equation}
where $U(y)$ is a local-$G$ element.  Then the original first solution
\begin{equation}
\chi = \chi_1(y),\qquad A_M = 0
\end{equation}
is gauge-equivalent to
\begin{equation}
\chi = \chi_2(y),\qquad A_M = -\frac{i}{e}(\partial_M U) U^{\dagger},
\label{eq:localxfm}
\end{equation}
where $e$ is the gauge coupling constant, but it is {\em not} gauge-equivalent to
\begin{equation}
\chi = \chi_2(y),\qquad A_M = 0,
\end{equation}
which is the original second solution.  Thus the two solutions $\chi = \chi_1,\ A_M = 0$
and $\chi = \chi_2,\ A_M = 0$ have different energy densities, even though the scalar-field
portions are related by a local symmetry transformation.  Although 
$A_M = -\frac{i}{e}(\partial_M U) U^{\dagger}$ is a pure-gauge configuration, it contributes
to the energy density through the $\chi-A_M$ interaction terms.

Setting the gauge fields to zero at the solution-level is basically a convenient choice of
gauge, one we shall adopt from now on.  Of course the solutions can be made to look very
different by gauge-transforming them, but their physical consequences cannot change.  This
circumstance is no different from the monopole or local-string cases, where again the
solutions look different in different gauges.  Actually, it is no more complicated than the
usual homogeneous vacuum expectation value (VEV) case.  If $\langle\chi\rangle$ is a homogeneous
VEV, then it can be gauge-transformed to a non-homogeneous configuration $U(x) \langle\chi\rangle$
but the scalar gradient energy is cancelled by the gauge-field contribution.

The alert reader may have noticed the following:  we have not proven that the minimum-energy
DW configuration must have a gauge-field sector that is gauge-equivalent to zero.  This does
indeed appear to be a loose end.  We shall make the assumption that it is in fact true for
the purposes of the rest of this paper.  Ultimately, one could uncover its hypothetical
falsity by a perturbative stability analysis for the DW, but that is well beyond the scope
of the present investigation.

Finally, a technical point: The set of $H_g$ contains an uncountable infinity of 
differently-embedded but isomorphic subgroups. However, there is a certain useful
sense in which the number of embeddings can be considered finite.  Let the
Cartan subalgebra ${\cal G}_C$ of $G$ be a certain particular set of generators, corresponding to
a particular choice of basis for the Lie algebra.  If we require that 
the Cartan subalgebras of two subgroups $H_{g_1}$ and $H_{g_2}$ are both subspaces of ${\cal G}_C$,
then the number of distinct embeddings is finite.  A
familiar example of this concerns the $SU(2)$ subgroups of $SU(3)$.  While there are an
uncountable infinity of ways of embedding $SU(2)$ in $SU(3)$, there are only three
embeddings that have the $SU(2)$ Cartan subalgebras as subspaces of the given 
Cartan-subalgebra space of $SU(3)$.  These are usually called I-spin, U-spin and V-spin.
When we say ``different embeddings'' below, this is what we shall mean.\footnote{Note that taking
linear combinations of Cartan generators to define different embeddings is in accord
with Dynkin's general theory of embeddings \cite{DynkinEmbedding}.  In that formalism, the embedding of
an algebra ${\cal H}$ into a simple or semi-simple algebra ${\cal G}$ is fully defined by
a mapping $F$ from the Cartan subalgebra of ${\cal H}$ into the Cartan subalgebra of ${\cal G}$,
as per $H_{\alpha} \to F(H_{\alpha}) = \sum_{a=1}^n F_{\alpha a} G_a$, where $H_{\alpha}$ ($\alpha = 
1,2,\ldots,m$) and $G_a$ ($a = 1,2,\ldots,n$) are the Cartan generators of ${\cal H}$ and ${\cal G}$,
respectively.  The matrix $(F_{\alpha a})$ is the defining matrix of the embedding, and two
embeddings are different if their defining matrices are different.}

\section{$SO(10)$ warm-up example: the need for $E_6$}
\label{SO10}

We now discuss a $G = SO(10)$ model that serves both as a warm-up for $E_6$ and
explains why the extension to $E_6$ is necessary.  We shall make use of the $O(10)$-kink
analysis of Ref.~\cite{Shin}.  While some recapitulation is necessary for completeness, we shall
be as brief as possible, and the reader is referred to Ref.~\cite{Shin} for a detailed discussion.

Let $\chi$ be a scalar multiplet in the adjoint representation, the $45$, of $SO(10)$.
The most general quartic Higgs potential is
\begin{equation}
V = \frac{1}{2}\, \mu^2\, {\rm Tr}(\chi^2) + \frac{1}{4}\, \lambda_1\, 
{\rm Tr}(\chi^2)^2 + \frac{1}{4}\, \lambda_2\, {\rm Tr}(\chi^4),
\label{eq:V}
\end{equation}
where $\chi = f_{\alpha} \hat{X}^{\alpha}$ with the $\hat{X}$'s being matrix
representations of the generators in the fundamental of $SO(10)$ while the $f_{\alpha}$'s
are the components of the adjoint multiplet.  The matrix $\chi$ is
antisymmetric and transforms as
per $\chi \to A \chi A^T$ where $A$ is an $SO(10)$ fundamental-representation matrix.
The parameter $\mu^2$ is chosen to be positive since ${\rm Tr}(\chi^2)$ is negative definite.
The cubic invariant ${\rm Tr}(\chi^3)$ identically vanishes so there is an
accidental discrete $Z_2$ symmetry, $\chi \to -\chi$, which shall play the role of $Z$.
It is not a subgroup of $O(10)$.

The global minimisation of such a potential was performed by Li \cite{Li} (see also \cite{Kaymakcalan}).  
Using an $SO(10)$
transformation, one may always bring a VEV pattern into the standard form
\begin{equation}
\chi = {\rm diag}(f_1\, \epsilon\, ,\, f_2\, \epsilon\, ,\, f_3\, \epsilon\, ,\, f_4\, \epsilon\, ,\, f_5\, \epsilon),
\label{eq:standardform}
\end{equation}
where the $f_i$ are real and
\begin{equation}
\epsilon \equiv i\, \sigma_2 = \left( \begin{array}{cc}
0 & 1 \\ -1 & 0 \end{array} \right).
\end{equation}
The five independent fields $f_i$ correspond to the five generators in the $SO(10)$ Cartan
subalgebra.  In this basis,
\begin{equation}
V = -\mu^2\, \sum_{i=1}^{5} f_i^2 + \lambda_1\, \left( \sum_{i=1}^{5} f_i^2 \right)^2
+ \frac{1}{2}\, \lambda_2\, \sum_{i=1}^{5} f_i^4.
\label{eq:rewrittenV}
\end{equation}
For $\lambda_2 > 0$, the global minima of $V$ are at
\begin{equation}
f_i^2 = \frac{\mu^2}{10\lambda_1 + \lambda_2}\quad \equiv f^2_{\rm min}\ \forall i,
\label{eq:globalminU5}
\end{equation}
where we define $f_{\rm min} \equiv \sqrt{\mu^2/(10\lambda_1 + \lambda_2)}$
and the unbroken subgroup is $H = U(5)$.  The values of $f_i$ at the minima are 
specified up to a sign that can be chosen independently for each component
\begin{equation}
f_i = \pm f_{\rm min}.
\end{equation}
Different choices for these signs correspond to two features: different embeddings of $U(5)$ in
$SO(10)$ and also a choice of which $Z_2$ sector the minimum lies in.

To explore this further, let us turn to possible domain wall configurations.  Suppose
that at $y = -\infty$, we choose as our boundary condition
\begin{equation}
\chi(-\infty) = - f_{\rm min}^{(5)} \equiv - f_{\rm min}\, {\rm diag}(\epsilon\, ,\, \epsilon\, ,\, 
\epsilon\, ,\, \epsilon\, ,\, \epsilon).
\label{eq:Phivac1}
\end{equation}
This defines a certain $U(5)$ unbroken at $y = -\infty$, and the VEV lies in one of the two
disconnected pieces of the vacuum manifold.  At $y = +\infty$, there are three
choices that lie in the other piece of the vacuum manifold, disconnected from the first
by the spontaneously broken $Z_2$:
\begin{equation}
\chi(+\infty) = \left\{  \begin{array}{ccc}
f_{\rm min}^{(5)} & \equiv & f_{\rm min}\, {\rm diag}
(\epsilon\, ,\, \epsilon\, ,\, \epsilon\, ,\, \epsilon\, ,\, \epsilon) \\
f_{\rm min}^{(3,2)} & \equiv &  f_{\rm min}\, {\rm diag}
(\epsilon\, ,\, \epsilon\, ,\, \epsilon\, ,\, - \epsilon\, ,\, - \epsilon) \\
f_{\rm min}^{(4,1)} & \equiv & f_{\rm min}\, {\rm diag}
(\epsilon\, ,\, - \epsilon\, ,\, - \epsilon\, ,\, - \epsilon\, ,\, - \epsilon)
\end{array} \right. .
\label{eq:plusbc}
\end{equation}
(Permutations of the minus signs in the last two of these vacua are just a trivial
rearrangement of the representation-space and need not be separately considered.)
Vacua with an odd number of minus signs on the right-hand side on Eq.~(\ref{eq:plusbc})
are continuously connected to $\chi(-\infty)$ by $SO(10)$ and shall not be considered
as they would give rise to non-topological domain walls.

The three vacua in Eq.~(\ref{eq:plusbc}) are invariant under differently-embedded 
subgroups of $SO(10)$: $U(5)_1$, $U(5)_2$ and $U(5)_3$.
The superscripts $(5)$, $(3,2)$ and $(4,1)$ denote the numbers of plus
and minus signs in the VEVs.  But they also usefully describe
the unbroken symmetry of the domain wall at finite $y$, respectively
\begin{equation}
U(5),\qquad U(3)\otimes U(2)\quad {\rm and}\qquad U(4)\otimes U(1),
\end{equation}
as we now explain.

The ansatz for domain wall configurations that interpolate between the stated boundary 
conditions is $\chi(y) = h(y) \chi(-\infty) + g(y) \chi(+\infty)$, where the functions
$h$ and $g$ obey self-evident boundary conditions.
The first configuration, which interpolates between $-f_{\rm min}$ and $+f_{\rm min}$
{\rm for all} components $f_i(y)$, breaks $SO(10)$ to $U(5)_1$ at all values of $y$,
because the relative magnitudes of the components are always the same at a given y.  It
is a non-CoS domain wall.
The second configuration has an equal-magnitude $3 \times 3$ block (of $2 \times 2$ submatrices),
and an equal-magnitude $2 \times 2$ block.  The unbroken symmetry is then 
\begin{equation}
U(3) \otimes U(2) = U(5)_1 \cap U(5)_2.
\end{equation}
Similarly, the third configuration's block structure leads to $U(4) \otimes U(1) = U(5)_1 \cap U(5)_3$.

The Euler-Lagrange equations 
\begin{equation}
f_i'' = 2 \left[ -\mu^2 + 2 \lambda_1 \sum_{j=1}^{5} f_j^2 \right] f_i 
+ 2 \lambda_2 f_i^3,
\label{eq:generaldes}
\end{equation}
with the three types of boundary conditions above
may be solved numerically.  However, a simple way to prove that solutions exist is to consider the
$\lambda_1 = 0$ slice through parameter space.  The equations can then be solved analytically to
yield
\begin{equation}
f_i(y) = f_{\rm min} \tanh(\mu y)\ \forall i
\label{eq:5soln}
\end{equation}
for the first boundary condition choice,
\begin{equation}
f_i(y) = f_{\rm min} \tanh(\mu y)\ {\rm for}\ i=1,2,3\ \ {\rm and}\ \ f_i(y) = f_{\rm min}\ 
{\rm for}\ i=4,5
\label{eq:32soln}
\end{equation}
for the second choice, and
\begin{equation}
f_i(y) = f_{\rm min} \tanh(\mu y)\ {\rm for}\ i=1\ \ {\rm and}\ \ f_i(y) = f_{\rm min}\ 
{\rm for}\ i=2,3,4,5
\label{eq:41soln}
\end{equation}
for the third choice.  The surface energy densities are in the ratios $5:3:1$ for the first to
the third solutions \cite{Shin}, so Eq.~(\ref{eq:41soln}) gives the topologically stable configuration.

From a Dvali-Shifman point of view, this stable configuration has an unbroken $SU(4)$ on the
brane that is embedded in $SU(5)_1$ on the $y < 0$ side of the wall, and $SU(5)_3$ on the
$y > 0$ side.  The $SU(4)$ gauge bosons are thus localised to the wall, if the Dvali-Shifman
mechanism is correct, because by assumption both $SU(5)_1$ and $SU(5)_3$ are in confinement
phase in their respective bulk regions.  This establishes the connection between clash-of-symmetries
and Dvali-Shifman by way of an explicit rigorously worked-out solution.  It is, however, just a
toy model since the symmetry breaking pattern is not what is required phenomenologically.

The second configuration, with 
\begin{equation}
U(3) \otimes U(2) = SU(3) \otimes SU(2) \otimes U(1) \otimes U(1)
\label{eq:GSMxU1}
\end{equation}
on the brane is closer to what we need for a realistic model.  While the analytic solution of
Eq.~(\ref{eq:32soln}) is unstable to dynamical evolution to Eq.~(\ref{eq:41soln}), it could well
be that in another region of Higgs-potential parameter space the $U(3) \otimes U(2)$ solution
is the stable one.  This has not been established, but let us suppose it is true.  The model
then still does not quite work, although it comes close.

It is certainly true that the $SU(3) \otimes SU(2)$ factor in Eq.~(\ref{eq:GSMxU1}) is
Dvali-Shifman-localised, because it is a subgroup of both $SU(5)_1$ (the bulk symmetry
for $y < 0$) and $SU(5)_2$ (the bulk symmetry for $y > 0$).  However, there is a
problem with the hypercharge gauge boson corresponding to $U(1)_Y$.  To see this,
we need to examine the $U(1)$ generators more closely.

The asymptotic gauge groups are
\begin{equation}
U(5)_1 = SU(5)_1 \otimes U(1)_{X_1}\qquad {\rm and}\qquad U(5)_2 = SU(5)_2 \otimes U(1)_{X_2}.
\end{equation}
Denote by $Y_1$ the hypercharge generator inside $SU(5)_1$, and $Y_2$ the one inside
$SU(5)_2$.  The two $U(1)$'s in Eq.(\ref{eq:GSMxU1}) can be taken to be generated
either by $Y_1$ and $X_1$, or by $Y_2$ and $X_2$, and each pair can be written as
linear combinations of the other pair.  Now, either $Y_1$ or $Y_2$ can be the physical hypercharge $Y$.
Which one is selected will be an accident of spontaneous symmetry breaking.
At some scale above the electroweak, the breaking
\begin{equation}
U(1)_{Y_1} \otimes U(1)_{X_1} = U(1)_{Y_2} \otimes U(1)_{X_2} \to U(1)_Y,
\end{equation}
with either $Y = Y_1$ or $Y = Y_2$,
will have to take place to produce an effective standard model at low-energies (this will
require an additional Higgs field).  Suppose $Y = Y_1$ is spontaneously selected.
Then the hypercharge gauge boson cannot propagate into the $y < 0$ bulk, but a
component of it will propagate into the $y > 0$ bulk.  The generator $Y = Y_1$ is a linear
combination of $Y_2$ and $X_2$, so the hypercharge gauge field is a linear combination of
the gauge fields of $Y_2$ and $X_2$.  But only the $Y_2$ part is unable to propagate into
the $y > 0$ region; the $X_2$ part is immune from the Dvali-Shifman effect because it is
not confining.  After electroweak symmetry breaking, this will imply that both the photon
and $Z^0$ will leak into the $y > 0$ bulk, which is phenomenologically ruled out.  If $Y_2$
happens to become the physical $Y$, then leakage into $y < 0$ will occur.

This structure, with localised gluons and $W^{\pm}$ bosons, but semi-delocalised photons
and $Z^0$'s, {\rm almost} works.  But understanding its pathology also provides the cure:
We need to expand the symmetry on the brane to contain a full $SU(5)$, with the physical
hypercharge identified with one of its generators.  Further, this brane-$SU(5)$ must be a
subgroup of confining non-Abelian groups on both sides of the domain wall.  These
two features automatically arise when we upgrade from $SO(10)$ to $E_6$ as the symmetry of
the action.

\section{The $E_6$ domain-wall brane.}
\label{E6}

\subsection{Group theory}

Take a scalar field multiplet $\chi$ in the 78-dimensional adjoint representation of $E_6$.
In the next subsection, we shall analyse the associated Higgs potential and produce
domain wall solutions.  But for
now, we just need to use the fact that for a range of parameters the global minima of the
potential will induce
\begin{equation}
E_6 \to SO(10) \otimes U(1),
\end{equation}
which is a maximal subgroup.  Now consider different embeddings of $SO(10) \otimes U(1)$ 
in $E_6$.\footnote{We mean the finite
number of embeddings in the sense of the final paragraph of Sec.~\ref{CoS}.}  We shall show
below that there is a pair of embeddings, which we shall call simply
$SO(10) \otimes U(1)_E$ and $SO(10)' \otimes U(1)_{E'}$, that is of particular
interest for model-building.\footnote{The second embedding
has been used in unified model building \cite{BandoE6, BandoE6_2, AndersonBlazek, MaekawaE6}.}
The domain wall solution we shall find in the next subsection interpolates between $\chi$ VEVs that break
$E_6$ to these different but isomorphic subgroups on opposite sides of the wall.  The
symmetry on the wall is then
\begin{equation}
[\, SO(10) \otimes U(1)_E\, ] \cap [\, SO(10)' \otimes U(1)_{E'}\, ] = SU(5) \otimes U(1)_E \otimes U(1)_{E'},
\label{eq:E6clash}
\end{equation}
as we shall establish.   Since $SU(5) \subset SO(10) \cap SO(10)'$, the Dvali-Shifman mechanism
localises all the $SU(5)$ gauge bosons to the domain wall, including the photon and the $Z^0$.

Let us look at the group theory in more detail.  Under
\begin{equation}
E_6 \to SO(10) \otimes U(1)_E \to [\, SU(5) \otimes U(1)_X\, ] \otimes U(1)_E,
\end{equation}
the fundamental 27-dimensional representation of $E_6$ branches as per
\begin{eqnarray}
27 & \to & 1(4) + 10(-2) + 16(1) \nonumber\\
& \to & 1^{(0,4)} + [\, 5^{(2,-2)} + 5^{*(-2,-2)}\, ] + [\, 1^{(-5,1)} + 5^{*(3,1)} + 10^{(-1,1)}\, ].
\label{eq:27branching}
\end{eqnarray}
The notation for $SO(10) \otimes U(1)_E$ representations is $D(12 E)$, where $D$ is the dimension
of the $SO(10)$ multiplet, and the $U(1)_E$ generator has been normalised as per
\begin{equation}
{\rm Tr}_{27}(E^2) = \frac{1}{2}.
\label{eq:Enorm}
\end{equation}
(We use $12E$ to make the charges integers for convenience.)  The $SU(5) \otimes U(1)_X \otimes U(1)_E$
notation is $D^{(4\sqrt{15}X,12E)}$ with
\begin{equation}
{\rm Tr}_{27}(X^2) = \frac{1}{2},\qquad {\rm Tr}_{27}(EX) = 0.
\label{eq:Xnorm}
\end{equation}
The second embedding is revealed by considering the linear combinations
\begin{equation}
X' = -\frac{1}{4}(X + \sqrt{15}E),\qquad E' = \frac{1}{4}(-\sqrt{15}X + E)
\label{eq:XEXpEp}
\end{equation}
that correspond to
\begin{equation}
E_6 \to SO(10)' \otimes U(1)_{E'} \to [\, SU(5) \otimes U(1)_{X'}\, ] \otimes U(1)_{E'}.
\end{equation}
Rewriting the multiplets from $D^{(4\sqrt{15}X,12E)}$ notation to $D_{(4\sqrt{15}X',12E')}$ notation,
we see that
\begin{eqnarray}
1^{(0,4)} & = & 1_{(-5,1)} \nonumber\\
5^{(2,-2)} & = & 5_{(2,-2)} \nonumber\\
5^{*(-2,-2)} & = & 5^*_{(3,1)} \nonumber\\
1^{(-5,1)} & = & 1_{(0,4)} \nonumber\\
5^{*(3,1)} & = & 5^*_{(-2,-2)},\nonumber\\
10^{(-1,1)} & = & 10_{(-1,1)},
\end{eqnarray}
so the $5^*$'s flip roles as do the singlets.  Let us now redundantly denote
the multiplets through
\begin{equation}
D^{(4\sqrt{15}X,12E)}_{(4\sqrt{15}X',12E')}.
\end{equation}
The $10$ of $SO(10)'$ is
\begin{equation}
5^{(2,-2)}_{(2,-2)} \oplus 5^{*(3,1)}_{(-2.-2)}, 
\end{equation}
whereas the $10$ of the original $SO(10)$ was instead formed by 
\begin{equation}
5^{(2,-2)}_{(2,-2)} \oplus 5^{*(-2,-2)}_{(3,1)}.
\end{equation}
Similarly, the $16$ of $SO(10)'$ consists of 
\begin{equation}
1^{(0,4)}_{(-5,1)} \oplus 5^{*(-2,-2)}_{(3,1)} \oplus 10^{(-1,1)}_{(-1,1)},
\end{equation}
whereas the $16$ of the original $SO(10)$ consisted of
\begin{equation} 
1^{(-5,1)}_{(0,4)} \oplus 5^{*(3,1)}_{(-2,-2)} \oplus 10^{(-1,1)}_{(-1,1)}.
\end{equation}
The $SO(10)'$ singlet is $1^{(-5,1)}_{(0,4)}$, whereas the original $SO(10)$ singlet
is $1^{(0,4)}_{(-5,1)}$.

Because all higher-dimensional representations of $E_6$ are formed from products of $27$'s,
the feature that some $SU(5) \otimes U(1)^2$ submultiplets flip when $(X,E) \to (X',E')$
propagates to all irreducible $E_6$ representations.  The submultiplets can be packaged
in $SO(10) \otimes U(1)_E$ multiplets, or repackaged into $SO(10)' \otimes U(1)_{E'}$
multiplets.  This establishes, constructively, that the two embeddings exist, and that
Eq.~(\ref{eq:E6clash}) is true.\footnote{By considering additional Cartan generators beyond
$E$ and $X$, more embeddings of $SO(10)$ can be found.  This is discussed further in the appendix.}  
Note that the additional $U(1)$'s are there because
adjoint configurations cannot rank-reduce.

Let us repeat this exercise for the adjoint of $E_6$:
\begin{eqnarray}
78 & \to & 1(0) + 45(0) + 16(-3) + 16^*(3)
\label{eq:78toSO10}\\
& \to & 1^{(0,0)}_{(0,0)} \nonumber\\
& + & [\, 1^{(0,0)}_{(0,0)} + 10^{(4,0)}_{(-1,-3)} + 10^{*(-4,0)}_{(1,3)} + 24^{(0,0)}_{(0,0)}\, ]\nonumber\\
& + & [\, 1^{(-5,-3)}_{(5,3)} + 5^{*(3,-3)}_{(3,-3)} + 10^{(-1,-3)}_{(4,0)}\, ] \nonumber\\
& + & [\, 1^{(5,3)}_{(-5,-3)} + 5^{(-3,3)}_{(-3,3)} + 10^{*(1,3)}_{(-4,0)}\, ]
\label{eq:78toSU5U1U1}
\end{eqnarray}
The flipping of roles is evidently 
\begin{eqnarray}
1^{(0,0)}_{(0,0)} & \leftrightarrow & 1^{(0,0)}_{(0,0)},\nonumber\\
10^{*(-4,0)}_{(1,3)} & \leftrightarrow & 10^{*(1,3)}_{(-4,0)},\nonumber\\
10^{(4,0)}_{(-1,-3)} & \leftrightarrow & 10^{(-1,-3)}_{(4,0)},\nonumber\\
1^{(-5,-3)}_{(5,3)} & \leftrightarrow & 1^{(5,3)}_{(-5,-3)}.
\end{eqnarray}
The $SU(5)$ adjoint $24^{(0,0)}_{(0,0)}$ is common to both $SO(10)$ embeddings,
as befits its status of being in the intersection of the two.

The two $1^{(0,0)}_{(0,0)}$ multiplets play important roles.  Giving a VEV to
the $1(0)$ in Eq.~(\ref{eq:78toSO10}) breaks $E_6$ to $SO(10) \otimes U(1)_E$, 
while a VEV for the
second singlet in Eq.~(\ref{eq:78toSU5U1U1}) breaks $E_6$ to $SO(10)' \otimes U(1)_{E'}$.  
A clash-of-symmetries kink interpolates between these two VEVs imposed as boundary
conditions. At $|y| < \infty$, both $SU(5) \otimes U(1)^2$ singlet components of the $78$ have nonzero
values, and this is precisely why the configuration breaks $E_6$ to the
intersection of the two subgroups.  To analyse this further, we must consider the dynamics.

\subsection{Higgs potential and domain-wall solutions}

The adjoint scalar multiplet $\chi$ shall be represented by
\begin{equation}
\chi = f_{\alpha} \hat{X}^{\alpha},\ \ \alpha = 1,\ldots,78
\label{eq:chiexpansion}
\end{equation}
where $\hat{X}$'s are matrix representations of the generators for the $27$ of $E_6$,
and the $f$'s are the field components. It transforms according to
\begin{equation}
\chi \to U \chi U^{\dagger}
\label{eq:chitransform}
\end{equation}
where $U$ is group representation matrix for the $27$.
We shall only be concerned with two of the
seventy-eight fields: those associated with $(E,E')$, equivalently $(X,E)$ or $(X',E')$
depending on what basis we choose for the Lie algebra.

We thus specialise to
\begin{equation}
\chi = f_E E + f_X X \equiv \tilde{f}_E E + f_{E'} E'
\label{eq:chitruncated}
\end{equation}
with
\begin{equation}
\tilde{f}_E \equiv f_E + \frac{f_X}{\sqrt{15}},\qquad f_{E'} \equiv -\frac{4 f_X}{\sqrt{15}},
\end{equation}
according to Eq.~(\ref{eq:XEXpEp}).  The $(X,E)$ basis is the more convenient for solving
the Euler-Lagrange equations, because $E$ and $X$ are orthogonal as per Eq.~(\ref{eq:Xnorm}).
The $(E,E')$ basis, however, is the simplest one for thinking about the two embeddings.

The VEVs we want for the boundary conditions are
\begin{equation}
(\tilde{f}_E, f_{E'}) = v (1,0),
\label{eq:10VEV}
\end{equation}
which corresponds to $E_6 \to SO(10) \otimes U(1)_E$.  The other VEV is
\begin{equation}
(\tilde{f}_E, f_{E'}) = -v (0,1)
\label{eq:10pVEV}
\end{equation}
which gives $E_6 \to SO(10)' \otimes U(1)_{E'}$.  The relative minus sign between
Eqs.~(\ref{eq:10VEV}) and (\ref{eq:10pVEV}) comes from the breaking of a
\begin{equation}
\chi \to -\chi
\label{eq:chiZ2}
\end{equation}
discrete symmetry we shall impose on the Higgs potential, and it is {\em crucial} for two reasons.
First, the spontaneous $Z_2$ breaking will ensure that our domain walls are
topologically non-trivial.  Second, it leads to a remarkable outcome for fermion
zero-mode localisation, to be explained in the next subsection.

In terms of the $(X,E)$ basis, these same VEVs are
\begin{equation}
(f_X, f_E) = v(0,1)\quad {\rm and}\quad v \left(\frac{\sqrt{15}}{4},-\frac{1}{4}\right),
\label{eq:EXVEVs}
\end{equation}
respectively.

We now need to find a Higgs potential with these two VEVs as degenerate global minima.
The Higgs potential is constructed out of adjoint invariants, which according to
Eqs.~(\ref{eq:chitransform}) and (\ref{eq:chiexpansion}) are
\begin{equation}
I_n = {\rm Tr}(\chi^n) = {\rm Tr}(\hat{X}^{\alpha_1}\hat{X}^{\alpha_2} \cdots \hat{X}^{\alpha_n})
f_{\alpha_1} f_{\alpha_2} \cdots f_{\alpha_n}.
\end{equation}
They are simply the nth order Casimir invariants.  According to Refs.~\cite{Racah, HarveyE6}, the independent
invariants are
\begin{equation}
I_2,\ I_5,\ I_6,\ I_8,\ I_9,\ I_{12},
\end{equation}
which immediately has an interesting consequence: the fact that $I_{5,9}$ are nonzero
means that the discrete $Z_2$ of Eq.(\ref{eq:chiZ2}) is {\em not a subgroup} of $E_6$,
because imposing it eliminates the otherwise present odd-power invariants.

It is sensible to truncate the Higgs potential at order-six:
\begin{equation}
V = - \lambda_1 I_2 + \lambda_2 (I_2)^2 - 2304 \kappa I_6 + \frac{4}{3} \lambda_3 (I_2)^3
\label{eq:VE6}
\end{equation}
where some peculiar numbers and signs have been inserted for later convenience.  In the extra-dimensional
setting, field-theoretic models must generally be considered as effective 
theories valid below an ultraviolet cutoff scale $\Lambda$, because they are almost
inevitably non-renormalisable.  In writing down a Higgs potential, one simply adds terms of
ever higher mass-dimension and truncates appropriately, given that the higher the 
mass-dimension the more suppressed it should become.  For the $E_6$ application, it is not
helpful to truncate at fourth order, because the only fourth-order invariant is $(I_2)^2$
and $I_2$ is invariant under an accidental $O(78)$ symmetry.  The presence of $I_6$ reduces
the symmetry of the Higgs potential to $E_6$ (presumably), and eliminates a pseudo-Goldstone 
boson issue.\footnote{An alternative is to truncate the classical theory at fourth order, 
but to add a Coleman-Weinberg
potential generated through quantum corrections that explicitly break the $O(78)$ \cite{HarveyE6}.}  

Equation (\ref{eq:VE6}) is a complicated sextic in seventy-eight fields.  But to perform
the global minimisation analysis, one can always transform any VEV pattern to
a standard form given by linear combinations of just the six generators in the Cartan
subalgebra of $E_6$.  This produces a still quite complicated sextic in six fields.
To make our discussion as simple as possible, in the main body of the paper
we shall further truncate to just the two Cartan
subalgebra generators of interest, and use Eq.~(\ref{eq:chitruncated}).  We extend the
global minimisation analysis to all six fields in the appendix.

With just $f_{E,X} \neq 0$, the nth-order invariant simplifies to
\begin{equation}
I_n = \sum_{k=0}^{n} \left( \begin{array}{c} n \\ n-k \end{array} \right) 
{\rm Tr}(E^{n-k} X^k) f_E^{n-k} f_X^k.
\end{equation}
The traces can be worked out by hand, because we know the matrix representations of
$E$ and $X$ from the branching rules in Eq.~(\ref{eq:27branching}).  We obtain
\begin{eqnarray}
I_2 & = & \frac{1}{2} ( f_E^2 + f_X^2 ),\label{eq:I2}\\
I_6 & = & \frac{1}{2304}\left(f_E^6 + 5 f_E^4 f_X^2 + 7 f_E^2 f_X^4 - 
\frac{48}{5\sqrt{15}}f_E f_X^5 + \frac{83}{25} f_X^6\right).
\end{eqnarray}
To understand the extrema of Eq.~(\ref{eq:VE6}), it is helpful to use the polar
decomposition
\begin{equation}
f_E = r \cos\theta,\qquad f_X = r \sin\theta.
\end{equation}
The VEVs of Eq.~(\ref{eq:EXVEVs}) are then
\begin{eqnarray}
(10,+)&:&\qquad \theta  = 0, \label{eq:VEV1}\\
(10',-)&:&\qquad \cos\theta = -\frac{1}{4},\ \sin\theta = \frac{\sqrt{15}}{4}.
\label{eq:minusVEV1prime}
\end{eqnarray}
The notation $(10,+)$ means that the VEV of Eq.~(\ref{eq:VEV1}) 
induces $E_6 \to SO(10) \otimes U(1)_E$, and we have (arbitrarily)
assigned it a positive $Z_2$ ``parity'' which signals that it lies in $G/H$ rather than $(G/H)_z$. 
Similarly, $(10',-)$ means $E_6 \to SO(10)' \otimes U(1)_{E'}$ and it lies in $(G/H)_z$.
There is another pair, with the opposite $Z_2$ parities:
\begin{eqnarray}
(10,-)&:&\qquad \theta  =  \pi, \label{eq:minusVEV1}\\
(10',+)&:&\qquad \cos\theta = \frac{1}{4},\ \sin\theta = -\frac{\sqrt{15}}{4}.
\label{eq:VEV1prime}
\end{eqnarray}
The topological CoS domain wall connects $(10,+)$ and $(10',-)$, accompanied
by a CoS anti-domain-wall connecting $(10,-)$ and $(10',+)$.
The topological non-CoS domain walls connect $(10,+)$ with $(10,-)$ [breaking
$E_6$ to $SO(10) \otimes U(1)_E$ at all $y$], and $(10',+)$ with $(10',-)$ [breaking
$E_6$ to $SO(10)' \otimes U(1)_{E'}$ at all $y$].  There are also nontopological configurations:
(i) $(10,+)$ connected to $(10',+)$, and (ii) $(10,-)$ connected to $(10',+)$, which are both CoS-like.

\begin{figure} 
\includegraphics[width=0.46\textwidth]{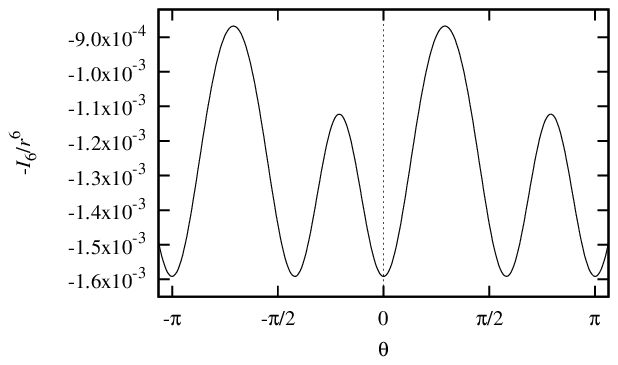}
\qquad \includegraphics[width=0.46\textwidth]{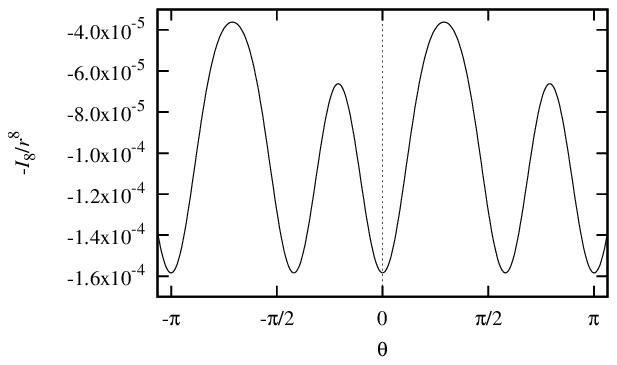}\\
\vspace{0.5cm}
\includegraphics[width=0.46\textwidth]{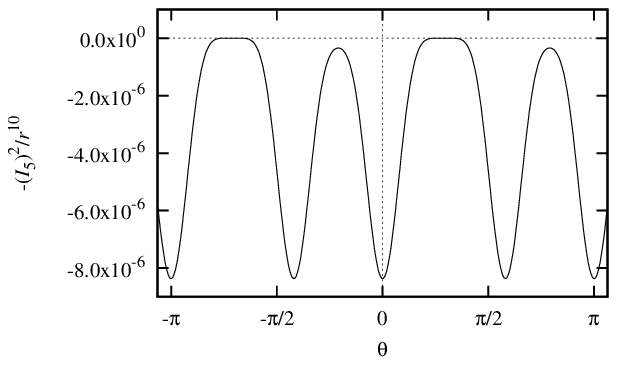}
\qquad \includegraphics[width=0.46\textwidth]{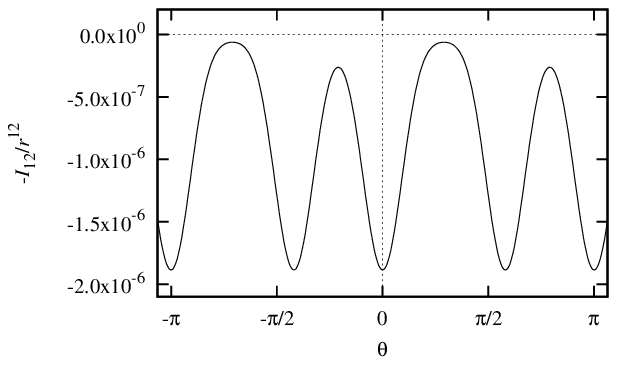}
  \caption{The invariants $-I_6/r^6$, $-I_8/r^8$, $-(I_5)^2/r^{10}$ and $-I_{12}/r^{12}$ as functions 
of $\theta$.  Note the remarkable similarity.  The global minima for all four,
reading from left to right, are: $(10,-)$, $(10',+)$, $(10,+)$, $(10',-)$ and then $(10,-)$ again
(see text for explanation of this notation).}
  \label{fig:invariants_angular}
\end{figure}

Figure \ref{fig:invariants_angular} display the invariants $-I_6/r^6$, $-I_8/r^8$, $-(I_5)^2/r^{10}$ and $-I_{12}/r^{12}$ as functions of $\theta$.  They show a remarkably similar structure.
It is evident that the global minima for all four invariants are precisely the four VEVs of Eqs.~(\ref{eq:VEV1}-\ref{eq:VEV1prime}).  It is clear from this that choosing to truncate at the
sextic level, as in Eq.~(\ref{eq:VE6}), does not sacrifice much in terms of generality.  We can be
confident that our simplified potential leads to solutions whose qualitative characteristics would
be retained were a wider class of higher-order potentials considered.  In addition, the appendix
shows that there are no deeper minima than $(10,\pm)$ and $(10',\pm)$ in the whole six-dimensional
Cartan domain.

\begin{figure}
\centering
\includegraphics[width=0.58\textwidth]{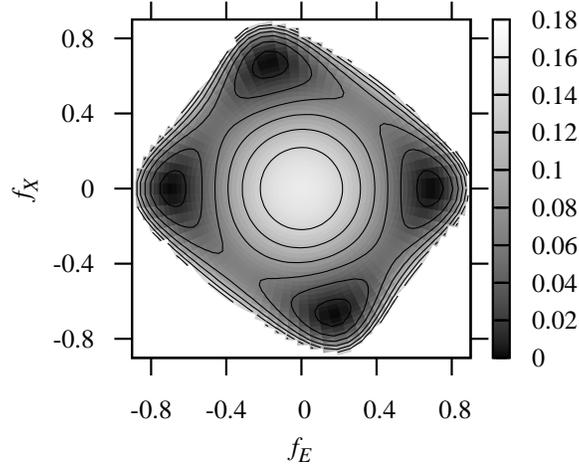}
  \caption{Contour plot of the
Higgs potential as a function of the two field components $f_E$ and $f_X$.
The parameters used are $\kappa=0.8$, $\lambda_1=1.0$, $\lambda_2 = 0$,
$\lambda_3=22.0$.  The darkest regions are the global minima in the order $(10,+)$, $(10',-)$,
$(10,-)$ and $(10',+)$ reading anticlockwise from the rightmost minimum.  The light area near the origin
is a local maximum.}
  \label{fig:higgs-pot}
\end{figure}

The sign in front of $I_6$ must be negative to achieve the desired extrema as minima rather than maxima.
The other terms in the Higgs potential, Eq.~(\ref{eq:VE6}), are independent of $\theta$, depending only on
the radial function $r$.  Hence, it is clear that Eq.~(\ref{eq:VE6}) has the global minima we require.
Figure~\ref{fig:higgs-pot} shows a contour plot of the Higgs potential for a certain
parameter choice illustrating this conclusion.  It is important
to realise that although the minima $(10,+)$ and $(10',+)$ [similarly
$(10,-)$ and $(10',-)$] look as though they are disconnected
by $E_6$, this is just an illusion created by only plotting the two-dimensional $(f_X,f_E)$ 
slice through the $78$-dimensional adjoint representation space.  Minima with opposite
parities are definitely disconnected from each other.\footnote{While it is certainly true that
the $Z_2$ is not a subgroup of $E_6$, so that in general $\chi$ and $-\chi$ are disconnected 
from each other, one may worry that there is nevertheless an $E_6$ transformation
that takes the specific configuration $(10,+)$ to $(10,-)$.  However, this is not the case.
Explicit calculation of the $E_6$ invariant $I_5$ reveals a nonzero $f_E^5$ term.  Hence, $f_E \to -f_E$
must be outside of $E_6$.  It does not matter that $I_5$ has been omitted from the Higgs potential,
as it is a purely group-theoretic argument.}

In the examples presented below, we further
simplify the Higgs potential by setting $\lambda_2 = 0$ as this term does not play an important role.
It is then easy to show that at the degenerate minima,
\begin{equation}
r = v \equiv \left( \frac{\lambda_1}{\lambda_3 - 22\kappa} \right)^{1/4},
\label{eq:v}
\end{equation}
so we must take $\lambda_3 > 22\kappa$, and that the value of $V$ at the minima is
\begin{equation}
V_{\rm min} = -\frac{1}{3} \sqrt{\frac{\lambda_1^3}{\lambda_3 - 22 \kappa}}.
\end{equation}
The latter must be subtracted from the potential
\begin{equation}
V \to V - V_{\rm min}
\end{equation}
to produce finite energy-densities for the domain wall configurations.  Figure~\ref{fig:higgs-pot}
is a contour plot of the potential energy showing the four degenerate global
minima.

\begin{figure}
\centering
\includegraphics[width=0.46\textwidth]{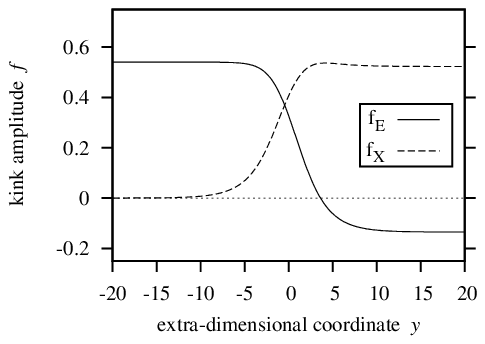}
\qquad
\includegraphics[width=0.46\textwidth]{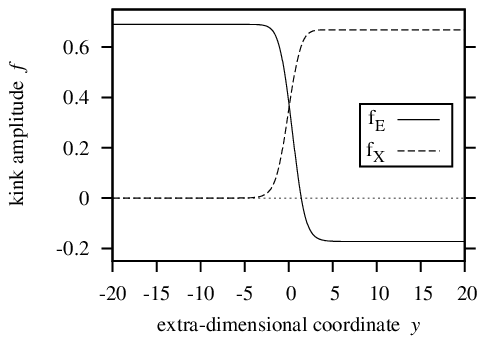}
  \caption{Clash-of-symmetries domain wall solutions interpolating 
between $(10,+)$ at $y = -\infty$ and $(10',-)$ at $y = +\infty$.
The parameters used in the left plot are $\kappa=0.2$, $\lambda_1=1.5$, $\lambda_2 = 0$, $\lambda_3=22.0$;
those in the right plot are $\kappa=0.8$, $\lambda_1=1.0$, $\lambda_2 = 0$, $\lambda_3=22.0$.}
  \label{fig:e6-kinks}
\end{figure}

Having understood the global minima, we may now solve the Euler-Lagrange equations
\begin{equation}
f''_X(y) = \frac{\partial V}{\partial f_X},\qquad f''_E(y) = \frac{\partial V}{\partial f_E}
\label{eq:E6EL}
\end{equation}
using those VEVs as boundary conditions.  Numerical solutions for CoS domain walls
interpolating between $(10,+)$ at $y = -\infty$ and $(10',-)$ at $y = +\infty$ with
two different parameter choices are displayed in Figure~\ref{fig:e6-kinks}.

\begin{figure}
\centering
\includegraphics[width=0.46\textwidth]{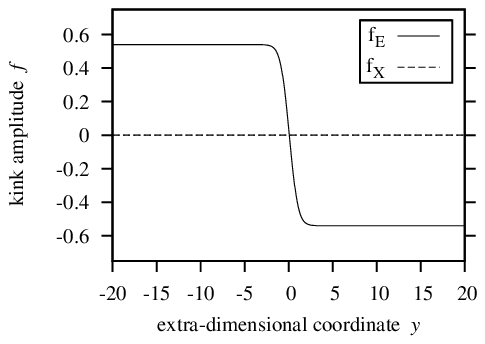}
\qquad
\includegraphics[width=0.46\textwidth]{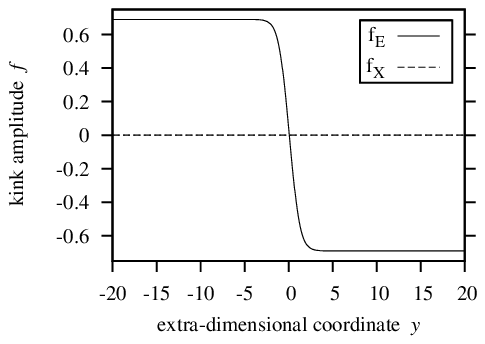}
  \caption{Non-clash-of-symmetries domain wall solutions interpolating 
between $(10,+)$ at $y = -\infty$ and $(10,-)$ at $y = +\infty$.
The parameters used in the left plot are $\kappa=0.2$, $\lambda_1=1.5$, $\lambda_2 = 0$, $\lambda_3=22.0$;
those in the right plot are $\kappa=0.8$, $\lambda_1=1.0$, $\lambda_2 = 0$, $\lambda_3=22.0$.}
  \label{fig:e6-non-kinks}
\end{figure}

\begin{figure}
\centering
\includegraphics[width=0.4\textwidth]{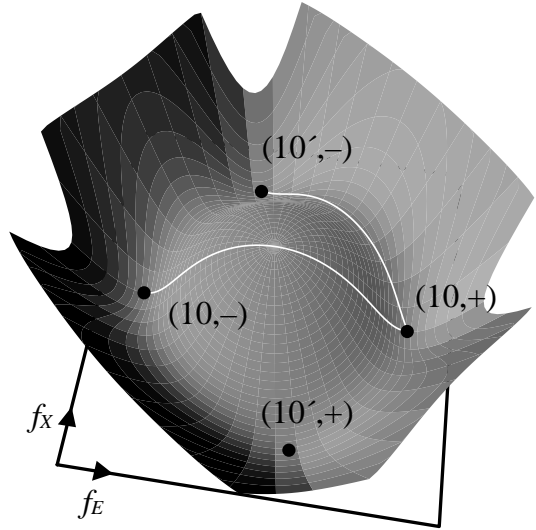}
  \caption{Three-dimensional plot of the
Higgs potential as a function of the two field components $f_E$ and $f_X$. The white lines
show the clash-of-symmetries domain wall (topmost) and the non-CoS
domain wall (bottommost).  The parameters used are $\kappa=0.8$, $\lambda_1=1.0$, $\lambda_2 = 0$,
$\lambda_3=22.0$.}
  \label{fig:e6pot3d}
\end{figure}

\begin{figure}
\centering
\includegraphics[width=0.4\textwidth]{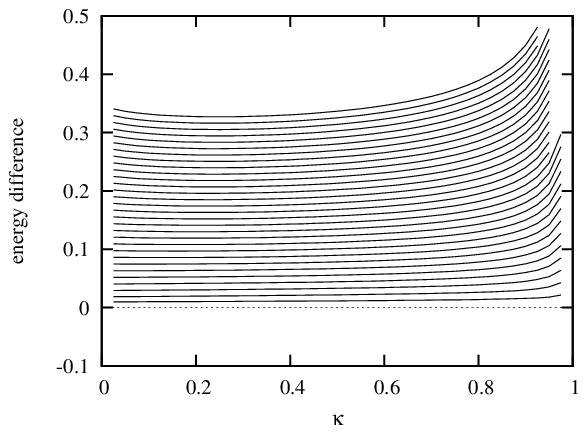}
  \caption{The difference in energy densities between the non-CoS and CoS domain wall solutions, 
$E_{\rm non-CoS} - E_{\rm CoS}$. 
We have numerically scanned through the parameter space with $0 < \kappa < 1$ along the horizontal 
axis, and each 
successive curve represents a different $\lambda_1$, beginning at $\lambda_1=0.05$ at the bottom and 
increasing in steps of $0.05$ to $\lambda_1=1.5$ at the top.  The energy difference is always positive, 
so the CoS domain wall has a lower energy.  We set $\lambda_2 = 0$ for simplicity.}
  \label{fig:energy-diff}
\end{figure}

Figure~\ref{fig:e6-non-kinks} depicts non-CoS domain wall solutions
for the same parameter choices.  The function
$f_X$ is zero, while $f_E$ interpolates between $v$ and $-v$ in archetypal kink fashion.  This means
that the non-CoS configurations feel the large potential-energy maximum at $f_X = f_E = 0$, while
the CoS configuration ``skirts around'' that central maximum.  This immediately implies that
the CoS solutions have lower energy density than the non-CoS solutions.  Although they are in the
same topological class, the CoS domain walls are stable while the non-CoS domain walls are unstable.
Figure~\ref{fig:e6pot3d} shows a three-dimensional plot of the potential and where the two DW configurations
sit with respect to the topography.  There is a tall maximum at the origin, 
and a corrugated valley encircling it, with four low points at the VEVs.
Figure \ref{fig:energy-diff} compares the energy densities of CoS and non-CoS domain walls.

\subsection{Fermion zero-mode localisation}

The CoS $E_6$ domain wall solutions described above are a good starting point for the
creation of domain-wall-brane models featuring $SU(5)$-invariant effective $3+1$-d theories
for localised fields.  To actually create such a model, fermions, additional Higgs bosons and 
gravitons have to be added.  In this subsection, we demonstrate that a phenomenologically-acceptable
fermion localisation pattern is obtained using the simplest possible mechanism.  We explain
why this is a remarkable result.

We simply Yukawa-couple a five-dimensional fermion multiplet in the $27$ of $E_6$,
\begin{equation}
\Psi \sim 27,
\end{equation}
to the adjoint scalar, as per
\begin{equation}
{\cal L}_Y = - h \overline{\Psi} \chi \Psi.
\end{equation}
We now substitute in the background CoS DW configuration for $\chi$ and solve the
resulting Dirac equations, which take the form
\begin{equation}
i \Gamma^M \partial_M \Psi^{(X,E)}(x^{\mu},y) - h [f_X(y) X + f_E(y) E]\Psi^{(X,E)}(x^{\mu},y) = 0.
\end{equation}
The notation $\Psi^{(X,E)}$ signifies the component of the $27$ with the specified $(X,E)$ charges,
as given in Eq.~(\ref{eq:27branching}).  The various components couple to different background
field configurations, 
\begin{equation}
b^{(X,E)}(y) = f_X(y) X + f_E(y) E,
\label{eq:bXE}
\end{equation}
given by the DW configuration and the charges.

The Dirac matrices are $\Gamma^M = (\gamma^\mu, -i\gamma_5)$.
We search for separated-variable solutions
\begin{equation}
\Psi(x^\mu,y) = F(y) \psi(x^\mu),
\end{equation}
demanding that $\psi$ have definite chirality, $\gamma_5 \psi = \pm\psi$, and obey the 
$3+1$-d massless Dirac equation, $i\gamma^{\mu} \partial_{\mu} \psi = 0$.  The 
solution for a profile is well known:
\begin{equation}
F^{(X,E)}(y) = N^{(X,E)} e^{- h \int^y b^{(X,E)}(y') dy'},
\end{equation}
where $N$ is a normalisation factor.   For the profile to represent localisation, it
must be square-integrable with respect to $y$.  For this to happen, $b^{(X,E)}$ must pass
through zero.  If so and it is an increasing function of $y$ (kink-like), then a left-(right-)handed
zero-mode occurs for $h>0(h<0)$.  If it passes through zero as a decreasing function (antikink-like),
then a left-(right-)handed zero-mode occurs for $h<0(h>0)$.

\begin{figure}
\centering
\includegraphics[width=0.48\textwidth]{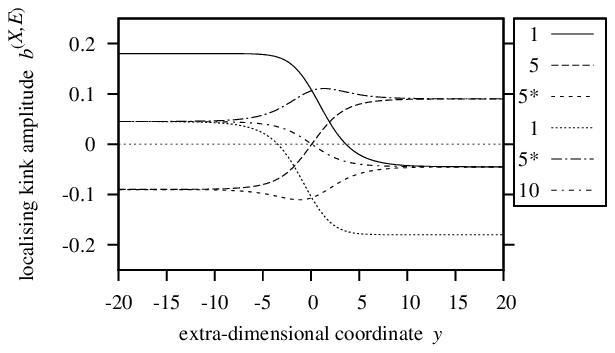}
\hfill
\includegraphics[width=0.48\textwidth]{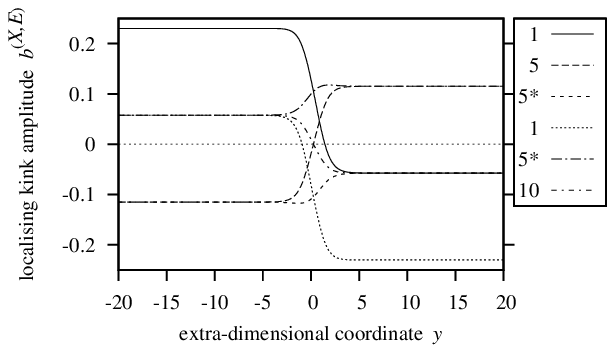}
  \caption{Clash-of-symmetries fermion localising profiles interpolating 
between $(10,+)$ at $y = -\infty$ and $(10',-)$ at $y = +\infty$.
The parameters used in the left plot are $\kappa=0.2$, $\lambda_1=1.5$, $\lambda_2 = 0$, $\lambda_3=22.0$;
those in the right plot are $\kappa=0.8$, $\lambda_1=1.0$, $\lambda_2 = 0$, $\lambda_3=22.0$.  The top to bottom
order of the $SU(5)$ fermion multiplets in the box on the right matches the order in Eq.~(\ref{eq:fermloc}).}
  \label{fig:e6-loc-profs}
\end{figure}

Figure~\ref{fig:e6-loc-profs} show the kink-like functions $b^{(X,E)}$ for the two parameter choices we have been
using as examples. Let us take $h$ to be negative:
\begin{equation}
h < 0.
\end{equation}
Using the notation $D^{(4\sqrt{15}X, 12E)}$ once again for the fermion
multiplets, the following displays these functions and states the localisation outcome,
which is either ``localised as left-handed (LH) zero-mode'' or ``localised as right-handed (RH) zero-mode''
or ``delocalised'':
\begin{eqnarray}
1^{(0,4)}:\ \frac{1}{3} f_E,\quad {\rm localised\ LH}
\nonumber\\
5^{(2,-2)}:\ \frac{1}{2} \left( \frac{1}{\sqrt{15}} f_X - \frac{1}{3} f_E \right),\quad {\rm localised\ RH}
\nonumber\\
5^{*(-2,-2)}:\ -\frac{1}{2} \left( \frac{1}{\sqrt{15}} f_X + \frac{1}{3} f_E \right)\,\quad {\rm delocalised}
\nonumber\\
1^{(-5,1)}:\ \frac{1}{4} \left( -\frac{5}{\sqrt{15}} f_X + \frac{1}{3} f_E \right),\quad {\rm localised\ LH}
\nonumber\\
5^{*(3,1)}:\  \frac{1}{4} \left( \frac{3}{\sqrt{15}} f_X + \frac{1}{3} f_E \right),\quad {\rm delocalised}
\nonumber\\
10^{(-1,1)}_{(-1,1)}:\ \frac{1}{4} \left( -\frac{1}{\sqrt{15}} f_X + \frac{1}{3} f_E \right),\quad {\rm localised\ LH}.
\label{eq:fermloc}
\end{eqnarray}
The two $5^*$'s are delocalised because the associated field never goes through zero.
The $5$ and the $10$ are localised at $y=0$ with opposite chiralities because their background fields are kink-like
and antikink-like, respectively.  The two singlets are localised at nonzero $y$ values, so the overall
spectrum is ``split''.

This is a remarkable outcome for two reasons.  First, because the $5$ is localised RH, it is 
equivalent to a LH-localised $5^*$. Thus the localised spectrum consists of LH zero-modes 
in the $SU(5)$ representation
\begin{equation}
5^* \oplus 10 \oplus 1 \oplus 1,
\end{equation}
in other words one standard family plus two singlet neutrinos.  Second, apart from the extra singlet, all the
exotic fermions in the $27$ of $E_6$ are delocalised and thus do not feature in the effective
$3+1$-d theory on the brane.  These benign outcomes depend crucially on the boundary condition choice
embodied by the CoS domain wall solution, including the $Z_2$ minus sign.  

Finally, there is an amusing aspect to this spectrum.  It resembles a usual $SO(10)$ family plus an
extra singlet.  However, the LH $5^*$, which is obtained from a $4+1$-d $5$, and the $10$ do not
come from a $16$ of either $SO(10)$ or $SO(10)'$.

\section{Conclusion}
\label{conc}

We find it extremely encouraging that, in the $E_6$ context, the clash-of-symmetries idea leads to
good outcomes for both gauge-boson localisation (assuming the
Dvali-Shifman mechanism works) and fermion localisation; that is the main point of this paper.

In summary, we have established a general connection between the clash-of-symmetries mechanism for
simultaneous brane-creation and internal-symmetry breaking with the Dvali-Shifman mechanism for
gauge boson localisation.  The two together provide a strong basis upon which to construct
realistic domain-wall-brane models.  These models should be compatible with type-2 Randall-Sundrum
graviton localisation (see Ref.~\cite{Volkasclash} for a CoS-style toy model featuring a 
background warped metric).

More specifically, we have found a domain wall solution in an $E_6$ adjoint-Higgs model that produces
an $SU(5) \otimes U(1)^2$ symmetry on the wall itself.  In one half of the bulk, the symmetry is 
enhanced to $SO(10) \otimes U(1)$, while in the other half of the bulk the enhancement is to
$SO(10)' \otimes U(1)'$.  The unprimed and primed groups are differently-embedded but isomorphic
subgroups of $E_6$.  Because the brane-$SU(5)$ is contained in both $SO(10)$ and $SO(10)'$,
the Dvali-Shifman localisation of its gauge bosons follows.  The simplest possible mechanism for
fermion zero-mode localisation produces a realistic spectrum, an outcome that depends on the
generic features of our domain wall configuration.

To complete a realistic model, one needs to add gravity (which is expected to be straightforward) and
to arrange for the additional spontaneous symmetry breaking cascade $SU(5) \otimes U(1)^2$ $\to$
$SU(3) \otimes SU(2) \otimes U(1)_Y$ $\to$ $SU(3) \otimes U(1)_Q$.  To achieve the latter, suitable
additional Higgs multiplets need to be introduced, and their background field configurations
have to be nonzero inside the domain wall to trigger the additional spontaneous symmetry breaking.
An example of this kind of dynamical structure is described in Ref.~\cite{DGV}, where the dominant background
domain-wall configuration breaks $SU(5)$ to $SU(3) \otimes SU(2) \otimes U(1)$, and then an additional Higgs
field induces electroweak symmetry breaking inside the wall.

\acknowledgments{RRV and AK were supported by the Australian Research Council, DPG by the Puzey Bequest
to the University of Melbourne, and KCW in part by funds provided by the U.S. Department of
Energy (D.O.E.) \#DE-FG02-92ER40702. AD is supported in part by the Albert Einstein Chair in Theoretical Physics.}

\appendix

\section{Full minimisation analysis}

We do this in two steps.  We first
extend the Higgs potential minimisation analysis by adding a third adjoint
component, associated with the Cartan sub-algebra generator identified as weak hypercharge $Y$.
This is useful because the result can be graphically visualised, and it reveals a third
embedding of $SO(10)$ that is related to the two embeddings used in the main body of the text.
In the second step, we report on a numerical study of the whole six-dimensional Cartan
subspace.

\begin{figure}
\centering
\includegraphics[width=0.7\textwidth]{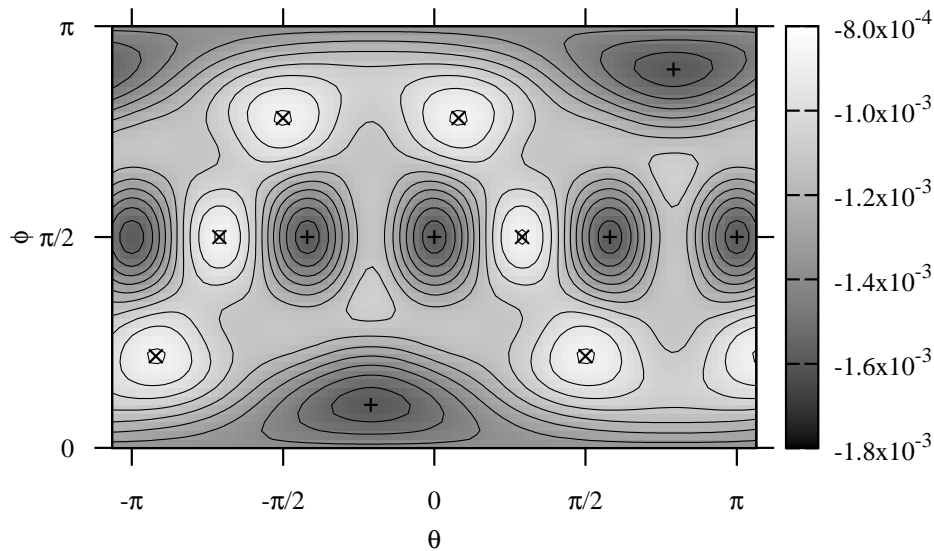}
  \caption{Contour plot of $-I_6/r^6$ as a function of $\theta$ and $\phi$. The degenerate global minima
breaking $E_6$ to various $SO(10) \otimes U(1)$ subgroups are marked with $+$ signs, while the
local maxima are indictaed with $\times$ signs.  The row of minima along $\phi = \pi/2$ correspond to
the global minima displayed in Fig.~\ref{fig:invariants_angular}.}
  \label{fig:i6contour}
\end{figure}

The truncated multiplet is thus first increased to
\begin{equation}
\chi = f_E E + f_X X + f_Y Y,
\end{equation}
where
\begin{eqnarray}
E & = & \frac{1}{12}{\rm diag}(4,-2,-2,-2,-2,-2,-2,-2,-2,-2,-2,1,1,1,1,1,1,1,1,1,1,1,1,1,1,1,1),\nonumber\\
X & = & \frac{1}{4\sqrt{15}}{\rm diag}(0,2,2,2,2,2,-2,-2,-2,-2,-2,-5,3,3,3,3,3,-1,-1,-1,-1,-1,-1,-1,-1,-1,-1),
\nonumber\\
Y & = & \frac{1}{2\sqrt{10}}{\rm diag}\left(0,1,1,-\frac{2}{3},-\frac{2}{3},-\frac{2}{3},-1,-1,\frac{2}{3},\frac{2}{3},0,-1,-1,
\frac{2}{3},\frac{2}{3},\frac{2}{3},\frac{1}{3},\frac{1}{3},\frac{1}{3},\frac{1}{3},\frac{1}{3},\frac{1}{3},
-\frac{4}{3},-\frac{4}{3},-\frac{4}{3},2 \right),
\end{eqnarray}
have been normalised as per
\begin{equation}
{\rm Tr}(E^2) = {\rm Tr}(X^2) = {\rm Tr}(Y^2) = 1/2,\qquad {\rm Tr}(EX) = {\rm Tr}(EY) = {\rm Tr}(XY) = 0.
\end{equation}
The sextic invariant is
\begin{equation}
I_6 = {\rm Tr}[(f_E E + f_X X + f_Y Y)^6].
\end{equation}
To visualise its structure, we go to a spherical-polar decomposition
\begin{equation}
f_E \equiv r \sin\phi \cos\theta,\ \ f_X \equiv r \sin\phi \sin\theta,\ \ f_Y \equiv r \cos\phi,
\end{equation}
which produces
\begin{eqnarray}
I_6 & = & \frac{r^6}{518400}\left(710 \cos^6\phi
+ 30 \cos^4\phi \left(51 + 4 \cos2\theta
-4 \sqrt{15} \sin2\theta\right) \sin^2\phi \right. \nonumber\\
& + &  60 \sqrt{2} \cos^3\phi \sin\theta \left(2 \sqrt{3}
+ 3 \sqrt{3} \cos2\theta
-\sqrt{5} \sin2\theta \right) \sin^3\phi \nonumber\\
& - & 45 \cos^2\phi \left(-34 + 2 \cos2\theta
+ 7 \cos4\theta - 2 \sqrt{15} \sin2\theta
+ \sqrt{15} \sin4\theta \right) \sin^4\phi \nonumber\\
& + & \frac{3}{2} \left(440 + 15 \cos2\theta
+ 84 \cos4\theta + 11 \cos6\theta
- 15 \sqrt{15} \sin2\theta
+ 12 \sqrt{15} \sin4\theta
- 3 \sqrt{15} \sin6\theta\right) \sin^6\phi \nonumber\\
& + & \left. 144 \cos^5\phi \left(\sqrt{10} \cos\theta \sin\phi
-\sqrt{6} \sin\theta  \sin\phi \right)\right)
\end{eqnarray}
Figure~\ref{fig:i6contour} plots $-I_6/r^6$ as a function of $\theta$ and $\phi$.
The $(E,X)$ plane is the line $\phi = \pi/2$, along which
the VEVs $(10,\pm)$ and $(10',\pm)$ can be seen.  Degenerate
with them are two more VEVs with $f_Y \neq 0$, located at
\begin{equation}
\theta = \arccos\left(-\frac{1}{2}\sqrt{\frac{5}{2}}\right),\quad \phi = \arccos\left(-\frac{3}{\sqrt{10}}\right)
\end{equation}
and
\begin{equation}
\theta = -\arccos\left(\frac{1}{2}\sqrt{\frac{5}{2}}\right),\quad \phi = \arccos\left(\frac{3}{\sqrt{10}}\right).
\end{equation}
The first VEV corresponds to a nonzero value for the adjoint component associated with
the generator $-E''$, where
\begin{equation}
E'' = \frac{1}{4} E - \frac{1}{4}\sqrt{\frac{3}{5}} X + \frac{3}{\sqrt{10}} Y.
\end{equation}
As the notation suggests, this minimum breaks $E_6$ to yet a third differently-embedded
subgroup which we can call $SO(10)'' \otimes U(1)_{E''}$, with negative $Z_2$ parity: $(10'',-)$.  
The second VEV is just $(10'',+)$.

The three groups $SO(10)$, $SO(10)'$ and $SO(10)''$ share a common $SU(3) \otimes SU(2)$
subgroup, but the $SU(5)$ contained in $SO(10) \cap SO(10)'$ is {\em not} a subgroup of $SO(10)''$ (this
is obvious, since $E''$ contains an admixture of $Y$ which is a generator of that $SU(5)$).
One can imagine a domain-wall junction configuration that utilises all three of these embeddings
for boundary conditions, but such a model would have a similar photon and $Z$-boson leakage
problem as the warm-up example of Sec.~\ref{SO10}.

The functions $-I_8/r^8$, $-(I_5)^2/r^{10}$ and $-I_{12}/r^{12}$ have exactly the same qualitative
structure as $-I_6/r^6$.  Thus $(10,\pm)$, $(10',\pm)$ and $(10'',\pm)$ will be the degenerate global
minima for a large class of potentials. The quadratic invariant is $\theta,\phi$-independent,
\begin{equation}
I_2 = \frac{1}{2} ( f_E^2 + f_X^2 + f_Y^2 )
\end{equation}
so one can simply add appropriate $(I_2)^n$ terms to the potential to ensure it is bounded from
below, and to generate a definite value for $r$ at the global minima.  The positions of the
global minima are determined entirely from the angular structure of the non-isotropic terms.

One can extend the analysis to all six Cartan components using a six-dimensional hyperspherical
polar decomposition.  The six fields are represented by one modulus, $r$, four zenith angles 
$0 \le \phi_{1,2,3,4} \le \pi$ and one azimuthal angle $-\pi \le \theta < \pi$.  Because the
group theoretic character of an extremum is determined entirely from the angular structure of the
invariants, a numerical study can readily be performed on the finite domain $(\phi_{1,2,3,4},\theta)$.
This study confirmed that the $E_6 \to SO(10) \otimes U(1)$ VEVs are the global minima of
$-I_{6,8,12}$ and $-(I_5)^2$.  The additional field dimensions simply revealed more degenerate vacua, corresponding to
extra embeddings of $SO(10)$ in $E_6$.  These new embeddings must correspond, physically
speaking, to choosing different $SU(3) \otimes SU(2)$ subgroups for colour and isospin.
The total number of $E_6 \to SO(10) \otimes U(1)$ extrema was found to be $54$, consisting of $27$ $Z_2$-related pairs.
This implies that, overall, there are $27$ embeddings of $SO(10) \otimes U(1)$ in $E_6$.  Though we
shall not display the results here, we have analytical expressions for the $27$ linear
combinations of Cartan generators that correspond to these VEVs.  In the breakdown
$27 \to 1 \oplus 10 \oplus 16$, these $27$ linear combinations turn out to be correlated
with the choice of which component to assign as the $SO(10)$ singlet in the decomposition.
A deeper reason for the number $27$ is perhaps the following: according to the $SU(3) \otimes SU(3) \otimes SU(3)$
maximal subgroup of $E_6$, there are three independent choices for the colour group.  The weak-isospin group
can then be selected as the $I-$, $U-$ or $V-$spin subgroup of either of the two remaining $SU(3)$'s. This
gives $3 \times 6 = 18$ choices for $SU(3) \otimes SU(2)$ embeddings.  According to our previous analysis, each
$SU(3) \otimes SU(2)$ is contained in the intersection of three different $SO(10)$'s, which suggests there
should be $18 \times 3 = 54$ embeddings of $SO(10)$.  However, recognising that $SO(10)$ contains an
$SU(2) \otimes SU(2)$ subgroup, we see that the correct number of independent embeddings is 
actually $54/2 = 27$.

\bibliography{references}

\end{document}